\documentclass[reprint,superscriptaddress,showpacs,amsmath,amssymb,aps,pra]{revtex4-1}
\usepackage{times}
\usepackage{graphicx}
\usepackage{amsthm}
\usepackage{mathrsfs}
\usepackage{subfigure}
\usepackage{hyperref}
\hypersetup{colorlinks=true, citecolor=blue, urlcolor=blue, linkcolor=blue}
\begin{document}
\title{Unified multivariate trace estimation and quantum error mitigation}
\author{Jin-Min Liang}
\email{jmliang@cnu.edu.cn}
\affiliation{School of Mathematical Sciences, Capital Normal University, Beijing 100048, China}
\author{Qiao-Qiao Lv}
\email{lvqq@cnu.edu.cn}
\affiliation{School of Mathematical Sciences, Capital Normal University, Beijing 100048, China}
\author{Zhi-Xi Wang}
\email{wangzhx@cnu.edu.cn}
\affiliation{School of Mathematical Sciences, Capital Normal University, Beijing 100048, China}
\author{Shao-Ming Fei}
\email{feishm@cnu.edu.cn}
\affiliation{School of Mathematical Sciences, Capital Normal University, Beijing 100048, China}
\date{\today}

\begin{abstract}
Calculating the trace of the product of $m$ $n$-qubit density matrices (multivariate trace) is a crucial subroutine in quantum error mitigation and information measures estimation. We propose an unified multivariate trace estimation (UMT) which conceptually unifies the previous qubit-optimal and depth-optimal approaches with tunable quantum circuit depth and the number of qubits. The constructed circuits have $\lceil(m-1)/s\rceil$ or $n\lceil(m-1)/s\rceil$ depth corresponding to $(s+m)n$ or $s+mn$ qubits for $s\in\{1,\cdots,\lfloor m/2\rfloor\}$, respectively. Such flexible circuit structures enable people to choose suitable circuits according different hardware devices. We apply UMT to virtual distillation for achieving exponential error suppression and design a family of concrete circuits to calculate the trace of the product of $8$ and $9$ $n$-qubit density matrices. Numerical example shows that the additional circuits still mitigate the noise expectation value under the global depolarizing channel.
\end{abstract}

\maketitle

\section{Introduction}
Fault-tolerant quantum (FTQ) computers may provide novel computational advantages over classical computers for many tasks \cite{Shor1994,Harrow2009Quantum,Biamonte2017Quantum,Liang2019Quantum}. As the perfect FTQ computers are not available yet, a preferable substitute is the the noisy intermediate-scale quantum (NISQ) devices with limited quantum resources \cite{Cerezo2021Variational,Bharti2022Noisy}. Nevertheless, the implementation of both FTQ and NISQ devices hinges on the effective control of noise. Thus, error correction and mitigation are of fundamental importance in quantum computation. Aiming at efficiently approximating the desired output states, the quantum error correction (QEC) provides a theoretical blueprint enabling quantum computation in an arbitrary small error level \cite{Devitt2013Quantum,Lidar2013Quantum}. However, duo to the larger qubit count, extra circuit complexity and other additional operations \cite{Preskill2018Quantum}, the overhead of QEC is too large to be available for practical applications. Therefore, a variety of quantum error mitigation (QEM) approaches \cite{Endo2021Hybrid,Zhang2021Variational,Bultrini2021Unifying,Yang2021Accelerated,Cai2021A,Takagi2022Universal,Huo2022Dual,Czarnik2022Improving} for NISQ algorithms have been presented instead for QEC. Different from QEC, QEM focuses on recovering the ideal measurement statistics (usually the expectation values) \cite{Cao2022NISQ} and can be directly employed in the ground state preparation \cite{Liang2022Improved,McArdle2019Variational}. For instance, the error extrapolation technique utilizes different error rates to the zero noise limit \cite{Temme2017Error,Li2017Efficient,Endo2018Practical}. In particular, probabilistically implementing the inverse process can mitigate the noise effect on computation for some noise channels \cite{Temme2017Error}. However, such error mitigation techniques rely on the prior knowledge on the noise model whose characterization is expensive.

The generalized quantum subspace expansion (GSE) \cite{Yoshioka2022Generalized} and the virtual distillation (VD) methods \cite{Koczor2021Exponential,Huggins2021Virtual} do not require any information of the noise. The GSE method optimizes states from a subspace expanded by a small subset of Pauli operators. It is effective against coherent errors \cite{Endo2021Hybrid}. The VD method prepares $m$ copies $n$-qubit noisy state $\rho=E_0|u_0\rangle\langle u_0|+\sum_{k=1}^{2^{n}-1}E_k|u_k\rangle\langle u_k|$
in spectral decomposition, $E_0>E_1\geq\cdots\geq E_{2^n-1}$, and calculates the expectation value of an observable $O$ with respect to the state $\rho^{m}/\textrm{Tr}(\rho^{m})$,
\begin{align}\label{VD}
\langle O\rangle_{\textrm{vd}}^{(m)}=\frac{\mbox{Tr}(O\rho^{m})}{\mbox{Tr}(\rho^{m})}=\langle O\rangle_{\textrm{exact}}[1+\mathcal{O}(E_1^m/E_0^m)],
\end{align}
which exponentially gets closer to the ideal value $\langle O\rangle_{\mbox{exact}}=\langle u_0|O|u_0\rangle$. The core idea behind this claim is that the state $\rho^{m}/\mbox{Tr}(\rho^{m})$ approaches to the desired pure state $|u_0\rangle$ exponentially in $m$. In order to obtain an exponential error suppression, an efficient quantum algorithm for measuring Eq. (\ref{VD}) is required.

Generally, instead of the $m$ copies of the $n$-qubit noisy state $\rho$, one may consider $m$ different $n$-qubit states $\rho_1,\cdots,\rho_m$.
The $\textrm{Tr}(\rho^{m})$ in Eq. (\ref{VD}) is then replaced by a more general quantity $\mbox{Tr}(\rho_1\cdots\rho_m)$ which is called multivariate trace (MT) first introduced in the work \cite{Quek2022Multivariate}. Measuring MT is of fundamental and practical interest in quantum information processing such as the calculation of the R\'{e}nyi entropy, entanglement entropy \cite{Yao2010Entanglement} and the entanglement spectroscopy \cite{Johri2017Entanglement}. Broadly speaking, the estimation of MT on a quantum computer is currently tackled by using either the qubit-optimal approach \cite{Ekert2002Direct} or depth-optimal method \cite{Quek2022Multivariate}. Here, the qubit-optimal means that the number of ancilla qubits is optimal, and the depth-optimal stands for that the circuit depth is minimal. The qubit-optimal approach requires only single ancilla qubit and a $\Theta(m)$-depth circuit, whereas the depth-optimal method requires $\lfloor m/2\rfloor$ ancilla qubits and a constant depth circuit, where $\lfloor\cdot\rfloor$ denotes the floor function. The former method is prohibited due to linear depth in $m$. Meanwhile, the latter method has an attractive depth for NISQ devices, but the linear number of needed qubits in $m$ would restrict its application to small $m$. Although a recent work \cite{Czarnik2021Qubit} drastically reduces the qubit resource by utilizing qubit resets technique, the circuit depth is still $\Theta(m)$. Thus, limited qubit number and circuit depth may hinder the advantage of the VD method in current NISQ era.

Accounting to the fact that the error accumulates with the increasing of the number of qubits and the depth of quantum circuit, in this work we provide a quantum algorithm to calculate the corrected expectation value Eq. (\ref{VD}) by constructing a family of circuits which have different circuit depth and number of qubits. As the authors \cite{Quek2022Multivariate} pointed out that their circuit is flexible and can be adjusted as different depth and number of qubits. Thus, in this work we first mathematically establish a specific trade-off relation between the number of qubits and the circuit depth. The circuit depth and the number of qubits can be denoted as a function of a free parameter $s$. From the variation  $s$, there are $\lfloor m/2\rfloor$ different circuit structures with the same number of quantum gates. Based on the constructed circuits, we propose an unified multivariate estimation (UMT), which is capable of calculating $\mbox{Tr}(\rho_1\cdots\rho_m)$ with a tunable circuit depth and number of qubits. The existing qubit-optimal and depth-optimal algorithms are two extremal cases of our algorithm for $s=1$ and $\lfloor m/2\rfloor$. Furthermore, we apply UMT to achieve the exponential error suppression and give a family of concrete circuits for $8$ and $9$ $n$-qubit density matrices. Finally, we simulate the effects of the global depolarizing channel in the process of estimating $\langle O\rangle_{\textrm{vd}}^{(5)}$ for a two qubits state $\rho$.

\section{Unified multivariate trace estimation}
The MT is defined as
\begin{align}\label{MT}
\mbox{Tr}(\rho_1\cdots\rho_{m}):=\mbox{Tr}[S^{(m)}(\rho_1\otimes\cdots\otimes\rho_m)],
\end{align}
where $S^{(m)}$ is a unitary representation of the cyclic shift permutation $\pi=(1,2,\cdots,m)$: $S^{(m)}|\psi_1\rangle\otimes|\psi_2\rangle\otimes\cdots\otimes|\psi_m\rangle
=|\psi_m\rangle\otimes|\psi_1\rangle\otimes\cdots\otimes|\psi_{m-1}\rangle$ for pure states $|\psi_1\rangle,\cdots,|\psi_m\rangle$. Note that for $m=2$, $\mbox{Tr}[\rho_1\rho_2]=\mbox{Tr}[S^{(2)}(\rho_1\otimes\rho_2)]$ with $S^{(2)}$ the SWAP operator. Eq. (\ref{MT}) means that the MT can be estimated by calculating the real and imaginary parts of $\mbox{Tr}[S^{(m)}(\rho_1\otimes\cdots\otimes\rho_m)]$. Following the framework of Ref. \cite{Ekert2002Direct}, a crucial step is to perform the controlled unitary $\mathcal{C}(S^{(m)})$ with respect to $S^{(m)}$, $\mathcal{C}(S^{(m)})=|0\rangle\langle0|\otimes I+|1\rangle\langle1|\otimes S^{(m)}$, where $I$ denotes the identity. In this section, we construct a sequence of alternative circuits to achieve the operation $\mathcal{C}(S^{(m)})$.

\subsection{The qubit-depth trade-off}

\newtheorem{proposition}{Proposition}[]
\begin{proposition}[Qubit-depth trade-off]\label{TradeOff1}
For a given set of $n$-qubit states $\{\rho_1,\cdots,\rho_m\}$, there exists a family of quantum circuits with depth $nh(m,s)$ and $(s+mn)$ qubits to achieve the operation $\mathcal{C}(S^{(m)})$, $s=1,\cdots,\lfloor m/2\rfloor$. The depth function $h(m,s)=\lceil(m-1)/s\rceil\in[2,m-1]$ is a monotonically decreasing function of the variable $s$.
\end{proposition}

The proposition \ref{TradeOff1} is based on the decomposition of the permutation cycle $\pi$ \cite{Quek2022Multivariate},
\begin{equation}
\pi=\left\{\begin{aligned}
& \prod_{j=2}^{m/2}(j,m+2-j)\prod_{i=1}^{m/2}(i,m+1-i),&m&~even\\
& \prod_{j=2}^{\lceil m/2\rceil}(j,m+2-j)\prod_{i=1}^{\lfloor m/2\rfloor}(i,m+1-i),&m&~odd
\end{aligned}\right.,\nonumber
\end{equation}
where all arithmetic is modulo $m$. $\pi$ can be decomposed into a product of $(m-1)$ transpositions. The circuit of proposition \ref{TradeOff1} contains an ancillary register (AR) and a work register (WR). The WR stores $m$ density matrices $\rho_1,\cdots,\rho_m$. The AR stores an $s$-qubit GHZ state, $|\mbox{GHZ}_s\rangle=\frac{1}{\sqrt{2}}(|0\rangle^{\otimes s}+|1\rangle^{\otimes s})$ which controls the SWAP operation between two different density matrices $\rho_l,\rho_k$ for $l,k=1,\cdots,m$. Each qubit of $|\mbox{GHZ}_s\rangle$ controls one transposition. Thus, the $s$-qubit GHZ state controls $s$ transpositions at one time. $m-1$ transpositions can be controlled at most $\lceil(m-1)/s\rceil$ times. Each SWAP operations can be decomposed into $n$ controlled SWAP gates. Thus, the total depth is $n\lceil(m-1)/s\rceil$. Here we remark that by using the qubit reset technique and the middle measurement, the $s$-qubit GHZ state $|\mbox{GHZ}_s\rangle$ can be prepared by a constant-depth quantum circuit (independent of $s$). The number of qubits needed is $s$ or $2(s-1)$ when $s$ is even or odd, respectively \cite{Quek2022Multivariate}. In general, for preparing $|\mbox{GHZ}_s\rangle$, a coherent quantum circuit has depth $\mathcal{O}(s)$. In this work, we do not consider the circuit depth of preparing state $|\mbox{GHZ}_s\rangle$ and density matrices $\rho_1,\cdots,\rho_m$.

\begin{figure}[ht]
\includegraphics[scale=0.4]{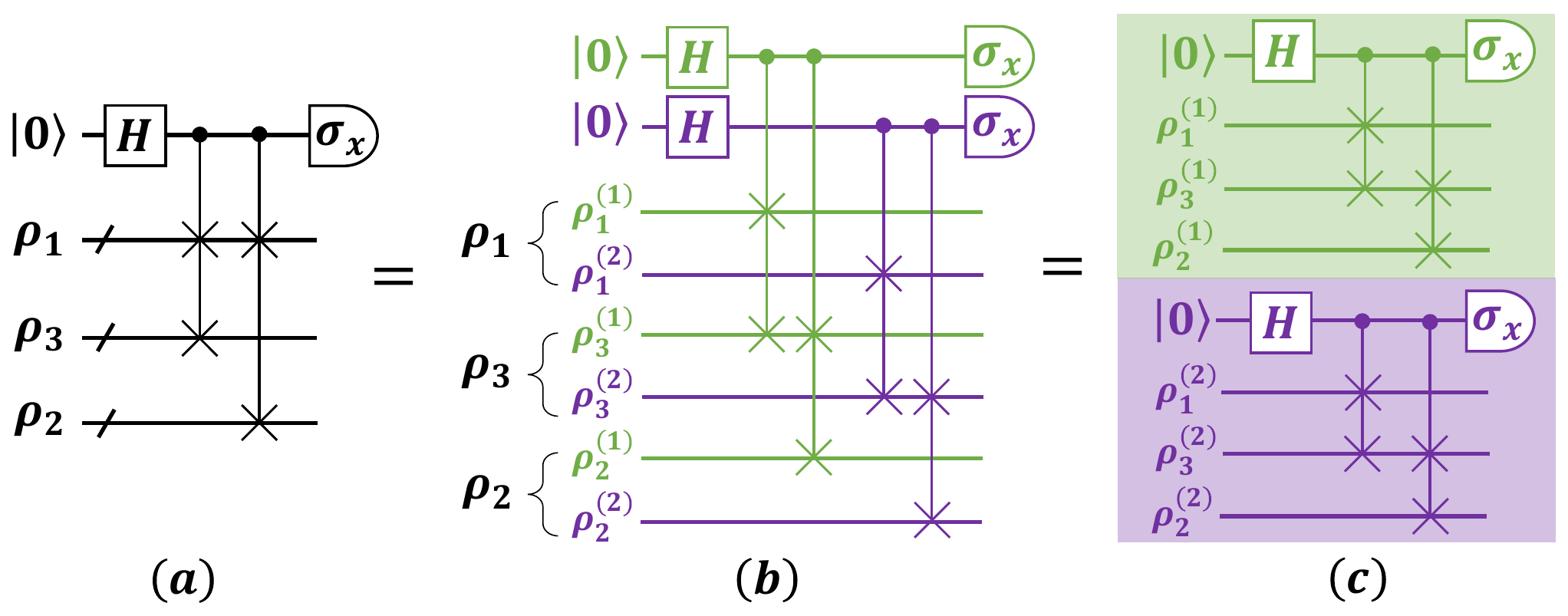}
\caption{Quantum circuits for estimating the real part of $\mbox{Tr}(\rho_1\rho_2\rho_3)$, where states $\rho_1,\rho_2,\rho_3$ are two-qubit ones. (a) The circuit in the work \cite{Ekert2002Direct}. The number of ancilla qubits is $1$ and the circuit depth is $4$. (b) and (c) are the parallelized versions of the circuit (a). The corresponding circuit depth is $2$ and the number of ancilla qubits is $2$. The $8$-qubit quantum circuit (b) can be seen as $2$ parallelized circuits with each $1$ ancilla qubit, as shown in (c).}
\label{Figure1}
\end{figure}

Notice that $s=1$ and $s=\lfloor m/2\rfloor$ cover the results of the qubit-optimal method \cite{Ekert2002Direct} and the depth-optimal approach \cite{Quek2022Multivariate}. With these two extremal cases, altogether there are $\lfloor m/2\rfloor$ optional circuits. In the work \cite{Ekert2002Direct} the authors introduced a single ancilla qubit and implemented a controlled unitary $S^{(m)}$ with depth $(m-1)n$. The circuit presented in \cite{Quek2022Multivariate} has depth $2n$ and $\lfloor m/2\rfloor$ ancilla qubits. The quantum circuits in \cite{Ekert2002Direct} and \cite{Quek2022Multivariate} can be seen as $n$ parallelized sub-circuits with only single or $\lfloor m/2\rfloor$ ancilla qubits, respectively \cite{Beckey2021Computable,Cai2021Resource}. Fig. \ref{Figure1} illustrates three mathematically equivalent quantum circuits for estimating $\mbox{Tr}(\rho_1\rho_2\rho_3)$ with single ancillary qubit.

Generalizing proposition \ref{TradeOff1} in the parallelized scenario, we have
\begin{proposition}[Qubit-depth trade-off in parallelized scenario]\label{TradeOff2}
Given a set of $n$-qubit states $\{\rho_1,\cdots,\rho_m\}$. There is a family of quantum circuits for achieving the operation $\mathcal{C}(S^{(m)})$. These circuits have depth $h(m,s)=\lceil(m-1)/s\rceil$ and $(s+m)n$ qubits, $s=1,\cdots,\lfloor m/2\rfloor$.
\end{proposition}

The circuits in proposition \ref{TradeOff2} are the parallelized versions of the Proposition \ref{TradeOff1}, in which the AR consists of $n$ $s$-qubit GHZ states. Each $s$-qubit GHZ state achieves the controlled SWAP operation on the single qubit subspace of the state $\rho_i$, $i=1,\cdots,m$. Moreover, in the case that $m=2$ and $\rho_1=\rho_2=\rho$, we recover the circuit for estimating the purity $\mbox{Tr}(\rho^2)$ presented in \cite{Beckey2021Computable} as a special case of our approach that $s=\lfloor m/2\rfloor=1$ and $h(2,1)=1$. In particular, when $s=2$, the circuit depth is $h(m,2)=m/2$ and $h(m,2)\in[\lfloor m/2\rfloor,\lceil m/2\rceil]$ for even and odd, respectively. This property guarantees that only $2n$ or $2$ additional ancilla qubits can reduce the depth by half. Proposition \ref{TradeOff1} and \ref{TradeOff2} show that the number of ancilla qubits and the circuit depth are tunable according to different hardware devices.

\subsection{UMT estimation}
Given the equivalent circuit construction of the controlled $S^{(m)}$ operation among density matrices, we perform a measurement in the Pauli $\sigma_x$ basis on each ancilla qubit. The expectation value gives an estimation of $\mbox{Tr}(\rho_1\cdots\rho_m)$. We have the following Theorem, see proof in Appendix \ref{T1Proof}.

\newtheorem{theorem}{Theorem}[]
\begin{theorem}[UMT estimation]\label{UMTEsti}
Given a set of $n$-qubit states $\{\rho_1,\cdots,\rho_m\}$ and fixed error $\varepsilon>0$ and $\delta\in(0,1)$. UMT estimation calculates the quantity Eq. (\ref{MT}) by the sample mean $\langle\hat{V}\rangle$ of a random variable $\hat{V}$ that can be produced using $\mathcal{O}(\varepsilon^{-2}\log(\delta^{-1}))$ repetitions of a quantum circuit, constructed via the Propositions \ref{TradeOff1} and \ref{TradeOff2}, consisting of $\mathcal{O}(mn)$ controlled SWAP gates such that
\begin{align}
\emph{Pr}(|\langle\hat{V}\rangle-\mbox{Tr}(\rho_1\cdots\rho_m)|\leq\varepsilon)\geq1-\delta
\end{align}
for $s=1,\cdots,\lfloor m/2\rfloor$.
\end{theorem}

Theorem \ref{UMTEsti} gives an analysis on the sample complexity guaranteed by the Hoeffding's inequality \cite{Hoeffding1963Probability}. The number of quantum gates is $\mathcal{O}(mn)$ for different circuit structures.

\section{UMT for virtual distillation}
A direct application of UMT is to the quantum error mitigation \cite{Huggins2021Virtual,Koczor2021Exponential}. Suppose that the near-term quantum devices aim to prepare an $n$-qubit pure state $|\phi\rangle$. However, owing to the effect of environment noise, one prepares instead a mixed state $\rho=\mathcal{C}(|\phi\rangle\langle\phi|)$, where the operation $\mathcal{C}$ is a map containing a unitary evolution and a noise channel such as depolarizing channel. The error-free expected value of an Hermitian operator $O$ is $\langle O\rangle=\mbox{Tr}(O|\phi\rangle\langle\phi|)$. However, the noisy expected value is $\langle O\rangle_{\textrm{noise}}=\mbox{Tr}(O\rho)\neq\langle O\rangle$. Virtual distillation provides a method to approximate $\langle O\rangle$ as a corrected expectation value
\begin{align}\label{ExpeValCor}
\langle O\rangle_{\textrm{vd}}^{(m)}=\frac{\mbox{Tr}(O\rho^{m})}{\mbox{Tr}(\rho^m)},
\end{align}
by $m$ copies of the mixed state $\rho$.

\subsection{Estimating the corrected expectation value with UMT}
It is clear to see that the denominator in Eq. (\ref{ExpeValCor}) can be evaluated by employing Theorem \ref{UMTEsti} with setting $\rho_1=\cdots=\rho_m=\rho$. The numerator of Eq. (\ref{ExpeValCor}) is
\begin{align}\label{Eq5}
\textrm{Tr}(O\rho^{m})=\textrm{Tr}(\tilde{O}^{(i)}S^{(m)}\rho\otimes\cdots\otimes\rho),
\end{align}
where the observable $\tilde{O}^{(i)}=I\otimes\cdots O^{(i)}\otimes\cdots\otimes I$ and $O^{(i)}$ denotes the operator $O$ acting on the $i$th register which stores the $i$th copies of $\rho$. Suppose an efficient decomposition
\begin{align}\label{ODecomposition}
O=\sum_{k=1}^{N_o}a_kP_k,~~a_k\in\mathbb{R},
\end{align}
where $P_k=\sigma_{k_1}\otimes\cdots\otimes\sigma_{k_n}$ are tensor products of Pauli operators $\sigma_{k_1},\cdots,\sigma_{k_n}\in\{\sigma_x,\sigma_y,\sigma_z,I\}$ and the quantity $\sum_{k=1}^{N_o}|a_k|=\mathcal{O}(c)$ is bounded by a constant $c$. It is straightforward to show that
\begin{align}
\textrm{Tr}(O\rho^{m})=\sum_{k=1}^{N_o}a_k\textrm{Tr}[\tilde{P}_k^{(i)}S^{(m)}
(\rho\otimes\cdots\otimes\rho)],
\end{align}
where $\tilde{P}_k=I\otimes\cdots P_k^{(i)}\otimes\cdots\otimes I$. By preparing $m$ copies of the state $\rho$, Theorem \ref{Enumerator} (see proof in Appendix \ref{T2Proof}) provides an estimator of the denominator.

\begin{theorem}[Estimation of $\mbox{Tr}(O\rho^{m})$]\label{Enumerator}
Let $\rho$ be an $n$-qubit noisy state. Given a Pauli decomposition of observable $O$ Eq. (\ref{ODecomposition}). For fixed precision $\varepsilon>0$, $\delta\in(0,1)$, and a constant $c$ there exists a quantum algorithm that estimates $\emph{Tr}(O\rho^{m})$ within $\varepsilon$ additive error with success probability $1-\delta$ and requires $\mathcal{O}\Big(\frac{mN_oc^2}{\varepsilon^2}\log\frac{1}{\delta}\Big)$ copies of $\rho$ and $\mathcal{O}\Big(\frac{N_oc^2}{\varepsilon^2}\log\frac{1}{\delta}\Big)$ repetitions of a quantum circuit (constructed via the propositions \ref{TradeOff1} and \ref{TradeOff2}) consisting of $\mathcal{O}(mn)$ controlled SWAP gates for $s=1,\cdots,\lfloor m/2\rfloor$.
\end{theorem}

After implementing the sequences of controlled SWAP gate, we perform a controlled $P_k$ on an arbitrary register storing the state $\rho$. Then we measure the ancilla qubits in the basis of Pauli operators $\sigma_x$ and $\sigma_y$. The measurement sample means are the real and imaginary parts of $\mbox{Tr}(O\rho^{m})$. We remark that the quantity $\sum_{k=1}^{N_o}|a_k|=c$ plays a great role in efficiently estimating the numerator $\mbox{Tr}(O\rho^{m})$. Due to the fact that the sample complexity is linear in $\left(\sum_{k=1}^{N_o}|a_k|\right)^2$, we thus expect that $\sum_{k=1}^{N_o}|a_k|$ is bounded by a constant $c$. This observation is intuitive. In variational quantum eigensolver \cite{Peruzzo2014Quantum} and variational quantum simulation \cite{Cerezo2021Variational,Bharti2022Noisy}, one typical question is the estimation of the expectation values of Hamiltonian $H$. The number of repetitions needed to obtain a precision $\epsilon$ with operator averaging is similar to our result \cite{Roggero2020Short}.

\subsection{Approximations for mean and variance of a Ratio}
The numerator and denominator are calculated via producing two independent variables $\hat{X}$ and $\hat{Y}$. Let $\hat{X}=(X_1,\cdots,X_{\mathcal{N}_O})$ and $\hat{Y}=(Y_1,\cdots,Y_{\mathcal{N}_I})$ be two independent variables, denoting the sampling results after running the UMT and measuring the ancilla qubits. The sample means are given by
\begin{align}
&\langle\hat{X}\rangle=\frac{\sum_{j=1}^{\mathcal{N}_O}X_j}
{\mathcal{N}_O}\approx\mathbb{E}[\hat{X}]=\textrm{Tr}(O\rho^{m}),\\
&\langle\hat{Y}\rangle=\frac{\sum_{j=1}^{\mathcal{N}_I}Y_j}
{\mathcal{N}_I}\approx\mathbb{E}[\hat{Y}]=\textrm{Tr}(\rho^{m}),
\end{align}
with error $\mathcal{E}_{O}=\mathcal{E}_{I}=\mathcal{O}(\mathcal{N}^{-1/2})$ such that
\begin{align}
&|\langle\hat{X}\rangle-\mathbb{E}[\hat{X}]|\leq\mathcal{E}_{O},~~
|\langle\hat{Y}\rangle-\mathbb{E}[\hat{Y}]|\leq\mathcal{E}_{I},
\end{align}
where we have set $\mathcal{N}_O=\mathcal{N}_I=\mathcal{N}$ to represent the number of samples. Then, the expectation value of the ratio $\langle\hat{X}\rangle/\langle\hat{Y}\rangle$ has an approximation, \cite{Small2010Expansions}
\begin{align}
\mathbb{E}\Bigg[\frac{\langle\hat{X}\rangle}{\langle\hat{Y}\rangle}\Bigg]
&\approx\frac{\mathbb{E}[\hat{X}]}{\mathbb{E}[\hat{Y}]}
+\frac{\mathbb{E}[\hat{X}]\textrm{Var}[\hat{Y}]^2}{\mathcal{N}\mathbb{E}[\hat{Y}]^3}\nonumber\\
&=\frac{\textrm{Tr}(O\rho^{m})}{\textrm{Tr}(\rho^{m})}
+\frac{\textrm{Tr}(O\rho^{m})\textrm{Var}[\hat{Y}]^2}{\mathcal{N}\textrm{Tr}(\rho^{m})^3}\nonumber\\
&=\frac{\textrm{Tr}(O\rho^{m})}{\textrm{Tr}(\rho^{m})}
+\frac{\textrm{Tr}(O\rho^{m})[1-\textrm{Tr}(\rho^{m})^2]^2}{\mathcal{N}\textrm{Tr}(\rho^{m})^3},
\end{align}
with error $\mathcal{O}(\mathcal{N}^{-1})$. The approximation variance of the ratio $\langle\hat{X}\rangle/\langle\hat{Y}\rangle$ \cite{Small2010Expansions} is
\begin{align}\label{variance}
\textrm{Var}\Bigg[\frac{\langle\hat{X}\rangle}{\langle\hat{Y}\rangle}\Bigg]
&\approx\frac{\mathbb{E}[\hat{X}]^2\textrm{Var}[\hat{Y}]^2}{\mathcal{N}\mathbb{E}[\hat{Y}]^4}
+\frac{\textrm{Var}[\hat{X}]^2}{\mathcal{N}\mathbb{E}[\hat{Y}]^2}\nonumber\\
&=\frac{\textrm{Tr}(O\rho^{m})^2\textrm{Var}[\hat{Y}]^2}{\mathcal{N}\textrm{Tr}(\rho^{m})^4}
+\frac{\textrm{Var}[\hat{X}]^2}{\mathcal{N}\textrm{Tr}(\rho^{m})^2}\nonumber\\
&=\frac{\textrm{Tr}(O\rho^{m})^2[1-\textrm{Tr}(\rho^{m})^2]^2}{\mathcal{N}\textrm{Tr}(\rho^{m})^4}\nonumber\\
&+\frac{\sum_{k=1}^{N_o}|a_k|^2[1-\textrm{Tr}(P_k\rho^{m})^2]}{\mathcal{N}\textrm{Tr}(\rho^{m})^2},
\end{align}
with error $\mathcal{O}(\mathcal{N}^{-1})$, where we have used the following results,
\begin{align}
&\textrm{Var}[\hat{X}]=\sum_{k=1}^{N_o}|a_k|^2[1-\textrm{Tr}(P_k\rho^{m})^2],\\
&\textrm{Var}[\hat{Y}]=1-\textrm{Tr}(\rho^{m})^2.
\end{align}
In particular, when $\rho$ is a pure state, the variance reduces to
\begin{align}
\textrm{Var}\Bigg[\frac{\langle\hat{X}\rangle}{\langle\hat{Y}\rangle}\Bigg]\approx
\frac{\sum_{k=1}^{N_o}|a_k|^2[1-\textrm{Tr}(P_k\rho^{m})^2]}{\mathcal{N}}.
\end{align}

The variance estimation provides an approach to evaluate the required number of samples. Assuming a desired variance is $\Delta^2$, Eq. (\ref{variance}) implies that the number of samples
\begin{align}
\mathcal{N}&\approx\frac{\textrm{Tr}(O\rho^{m})^2[1-\textrm{Tr}(\rho^{m})^2]^2}{\Delta^2\textrm{Tr}(\rho^{m})^4}\nonumber\\
&+\frac{\sum_{k=1}^{N_o}|a_k|^2[1-\textrm{Tr}(P_k\rho^{m})^2]}{\Delta^2\textrm{Tr}(\rho^{m})^2}.
\end{align}
In the work \cite{Huggins2021Virtual} the authors analyzed the variance of the estimator for $m=2$. Here, we present an approximation for the mean and variance of the estimator for arbitrary $m$.

\begin{figure}[ht]
\includegraphics[scale=0.4]{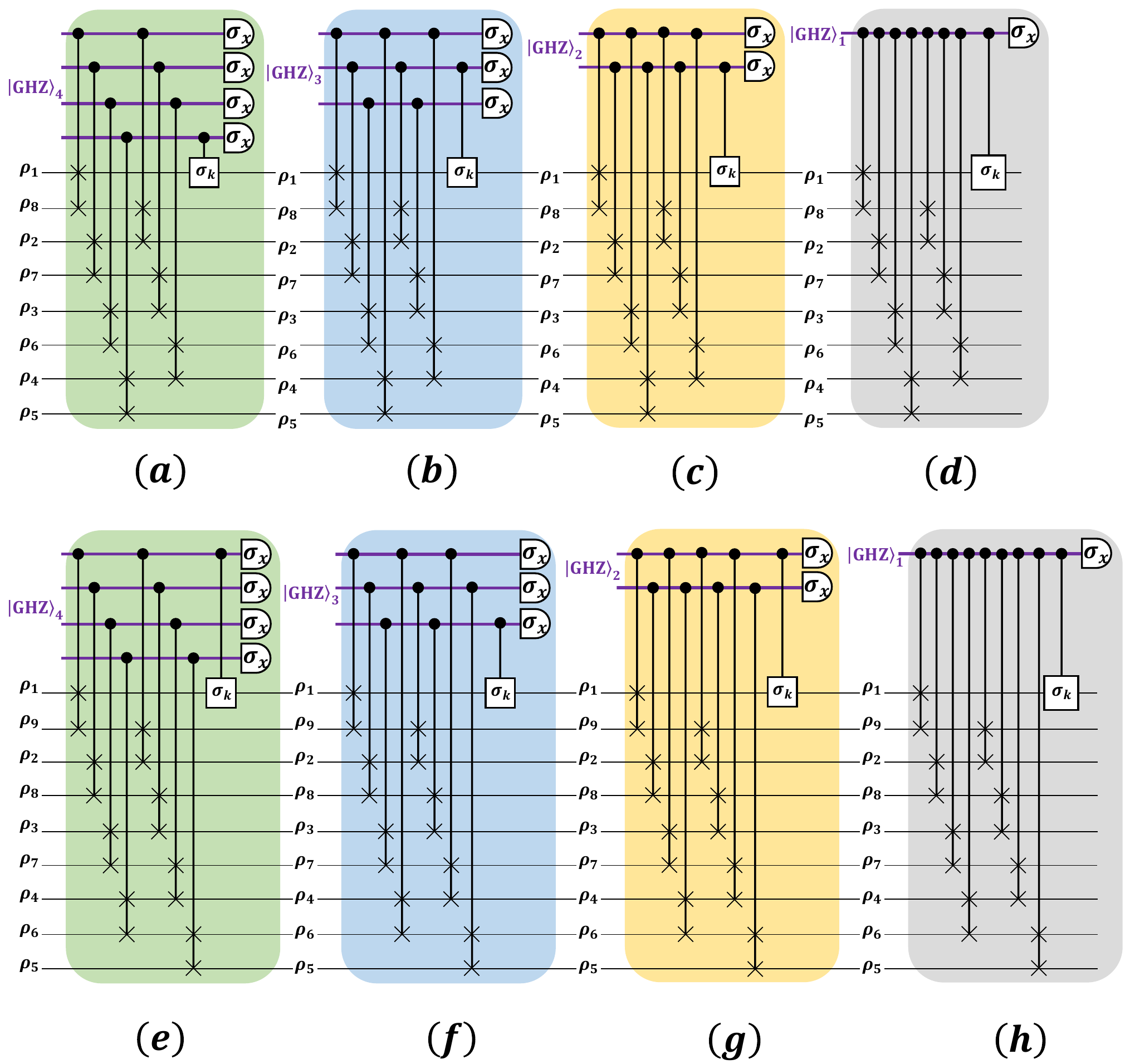}
\caption{Figures (a-d): The quantum circuits for estimating the real part of $\textrm{Tr}(O\rho^8)$ for $n$-qubit state $\rho$. Figures (e-h): The quantum circuits for estimating the real part of $\textrm{Tr}(O\rho^9)$ for $n$-qubit state $\rho$. Both circuits are from Proposition \ref{TradeOff1}. When setting $\sigma_k=I$, the circuits cover the real parts of $\textrm{Tr}(\rho^8)$ and $\textrm{Tr}(\rho^9)$. The imaginary parts of these quantities can be estimated by measuring the ancilla qubits in the basis of Pauli operator $\sigma_y$ and here $\rho_1=\cdots=\rho_8=\rho_9=\rho$.}
\label{Figure2}
\end{figure}

\begin{figure}[ht]
\includegraphics[scale=0.35]{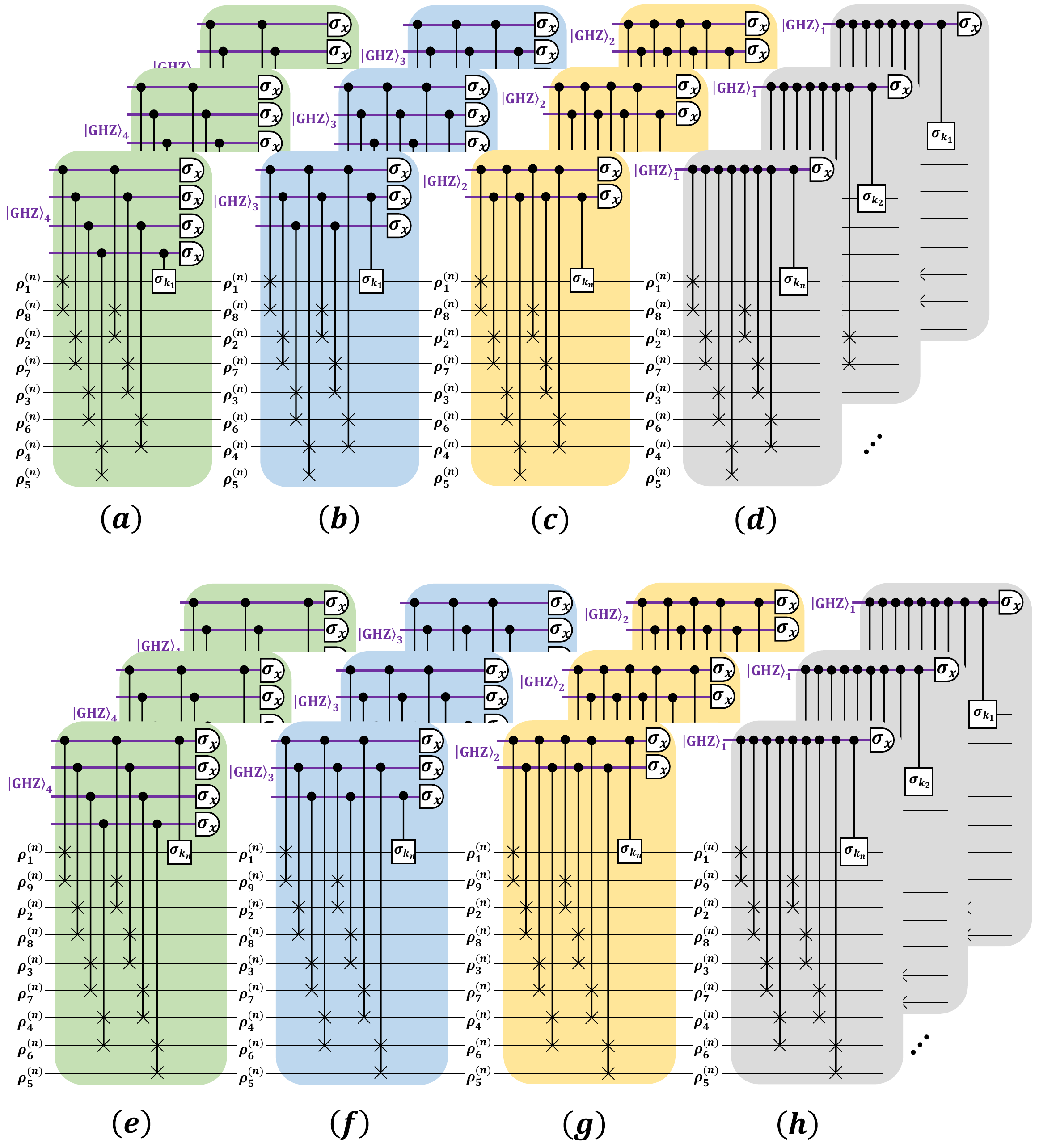}
\caption{Figures (a-d): The quantum circuits for estimating the real part of $\textrm{Tr}(O\rho^8)$ for $n$-qubit state $\rho$. (a) $s=4$, $h(8,4)=2$; (b) $s=3$, $h(8,3)=3$; (c) $s=2$, $h(8,2)=4$ and (d) $s=1$, $h(8,1)=7$. Figures (e-h): The quantum circuits for estimating the real part of $\textrm{Tr}(O\rho^9)$ for $n$-qubit state $\rho$. (e) $s=4$, $h(9,4)=2$; (f) $s=3$, $h(9,3)=4$; (g) $s=2$, $h(9,2)=4$ and (h) $s=1$, $h(9,1)=8$. Both circuits are given by Proposition \ref{TradeOff2}. When taking $\sigma_k=I$, the circuits cover the real parts of $\textrm{Tr}(\rho^8)$ and $\textrm{Tr}(\rho^9)$. The imaginary parts of these quantities can be estimated via the measurement on the ancilla qubits in the basis of Pauli operator $\sigma_y$.}
\label{Figure3}
\end{figure}

\subsection{Concrete construction of a family of circuits and noisy implementation}
In this section, we first show the circuit construction of $8$ and $9$ $n$-qubit density matrices. The permutations $\pi_8=(1,2,\cdots,8)$ and $\pi_9=(1,2,\cdots,9)$ have decompositions
\begin{align}
\pi_8&=(4,6)(3,7)(2,8)(4,5)(3,6)(2,7)(1,8),\\
\pi_9&=(5,6)(4,7)(3,8)(2,9)(4,6)(3,7)(2,8)(1,9),
\end{align}
where each of transpositions denotes the $n$ SWAP gates. Based on Proposition \ref{TradeOff1}, Fig. \ref{Figure2} (a-d) show $4$ circuits for computing $\textrm{Tr}(O\rho^8)$ and $\textrm{Tr}(\rho^8)$ for $s=1,2,3,4$. The total number of qubits is $s+8n$ including $s$ ancilla qubits. The depth is $2n$, $3n$, $4n$ and $7n$. Fig. \ref{Figure2} (e-h) show $4$ circuits for computing $\textrm{Tr}(O\rho^9)$ and $\textrm{Tr}(\rho^9)$ for $s=1,2,3,4$. The total number of qubits is $s+9n$ including $s$ ancilla qubits. The depth is $2n$, $3n$, $4n$ and $8n$. Fig. \ref{Figure3} is a parallelized version of Fig. \ref{Figure2} as shown in Proposition \ref{TradeOff2}. The total number of qubits is $(s+8)n$ including $sn$ ancilla qubits for $s=1,2,3,4$. The depth is $2$, $3$, $4$ and $7$, respectively.

\begin{figure}[ht]
\includegraphics[scale=0.6]{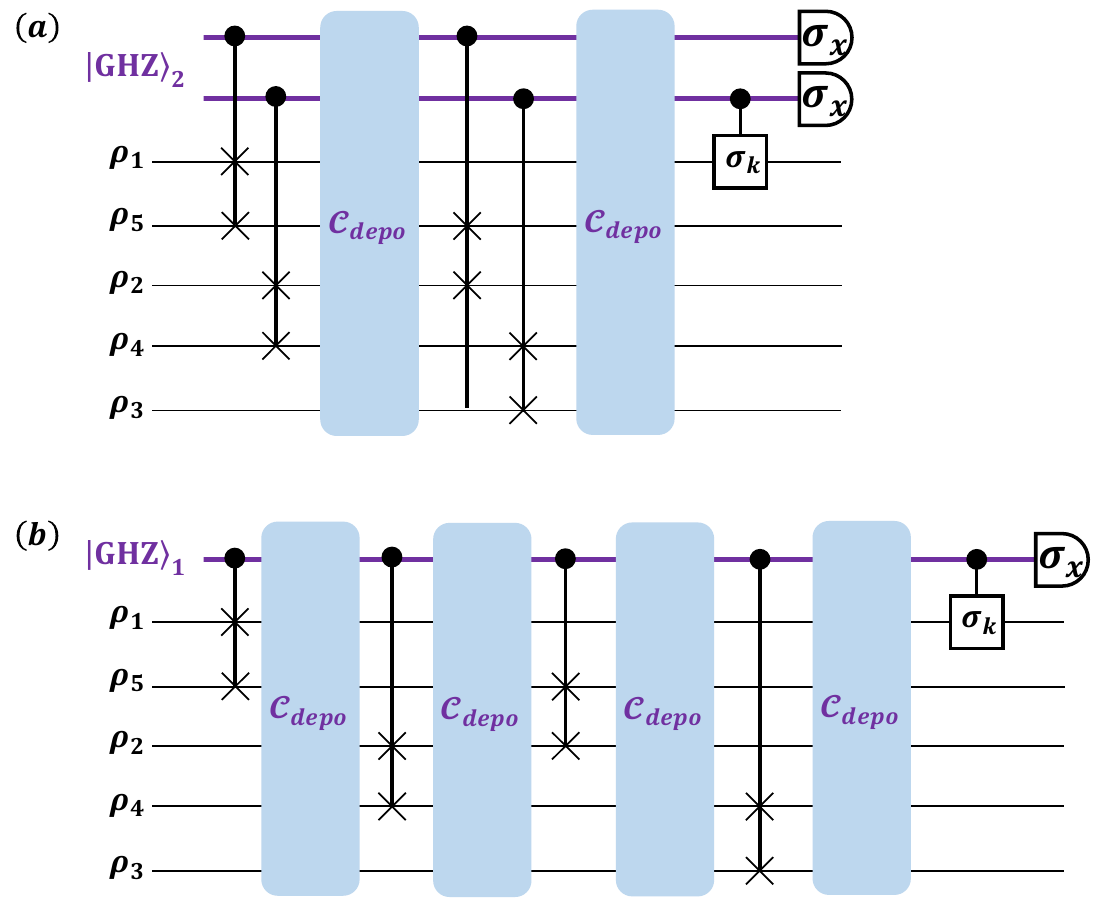}
\caption{The quantum circuits for estimating the real part of $\textrm{Tr}(O\rho_{\textrm{noise}}^5)$ for $2$-qubit state $\rho_{\textrm{noise}}$ with $\sigma_k=\sigma_z^{(1)}$ or $\sigma_z^{(2)}$, where $\rho_1=\cdots=\rho_5=\rho_{\textrm{noise}}$. (a) $s=2$, $h(5,2)=2$; (b) $s=1$, $h(5,1)=4$. When $\sigma_k=I$, the circuits cover the real part of $\textrm{Tr}(\rho_{\textrm{noise}}^5)$.}
\label{Figure4}
\end{figure}

Here, we consider the effect of noise in the quantum circuits for different width and depth. Suppose we prepare an exact state $\rho=U(\boldsymbol{\alpha})(|0\rangle\langle0|\otimes|0\rangle\langle0|)U^{\dag}(\boldsymbol{\alpha})$ by a parameterized quantum circuit
\begin{align}
U(\boldsymbol{\alpha})=\prod_{i=1}^{2}\mbox{CNOT}\times(R_y(\alpha_i)\otimes R_y(\alpha_{i+1})),
\end{align}
where $R_y(\alpha_i)=e^{-\iota\alpha_i/2\sigma_y}$ is the rotation operator, $\mbox{CNOT}$ denotes the controlled-NOT gate and the parameters $(\alpha_1,\alpha_2,\alpha_3,\alpha_4)=(0.8147,0.1270,0.2785,0.5469)$. The state $\rho$ after the depolarizing noise channel is given by
\begin{align}
\rho_{\textrm{noise}}=\mathcal{C}_{\textrm{depo}}(\gamma_0,\rho)=(1-\gamma_0)\rho+\gamma_0\frac{I}{4},~~~\gamma_0\in(0,1).
\end{align}
We measure the expectation value of observable $O=(\sigma_z^{(1)}+\sigma_z^{(2)})/2$ with respect to states $\rho$ and $\rho_{\textrm{noise}}$. Fig. \ref{Figure4} shows two circuits for estimating $\mbox{Tr}(O\rho_{\textrm{noise}}^5)$ and $\mbox{Tr}(\rho_{\textrm{noise}}^5)$. After each layer, we insert a depolarizing channel with parameter $\gamma$ for simulating the noise effects. Two and four depolarizing channels are required for Fig. \ref{Figure4}.(a) and (b), respectively. The ideal expectation value is $\langle O\rangle=\mbox{Tr}(O\rho)=0.7547$ and noise result $\langle O\rangle_{\textrm{noise}}=\mbox{Tr}(O\rho_{\textrm{noise}})=0.4528$ for $\gamma_0=0.4$. The corrected expectation value after VD and our circuits is given by
\begin{align}
\langle O\rangle_{\textrm{vd}}^{(5)}=\frac{\mbox{Tr}(O\rho_{\textrm{noise}}^5)}{\mbox{Tr}(\rho_{\textrm{noise}}^5)}.
\end{align}
As shown in Table \ref{Table1}, by numerical calculation it is shown that our circuits can still mitigate the error even in the presence of the depolarizing noise channel.
\begin{table}
\begin{tabular}{|c|c|c|c|c|}
\hline
 $\gamma$ & 0.2 & 0.4 & 0.6 & 0.8 \\ \hline
 Fig. \ref{Figure4}.(a) $\langle O\rangle_{\textrm{vd}}^{(5)}$ & 0.7546 & 0.7546 & 0.7546 & 0.7546 \\ \hline
 Fig. \ref{Figure4}.(b) $\langle O\rangle_{\textrm{vd}}^{(5)}$ & 0.7546 & 0.7546 & 0.7546 & 0.7546 \\ \hline
\end{tabular}
\caption{The corrected expectation value $\langle O\rangle_{\textrm{vd}}^{(5)}$ for different parameters $\gamma$.}
\label{Table1}
\end{table}

\section{Conclusion and discussion}
We have proposed an unified quantum algorithm for estimating the multivariate trace. Our results depend on a qubit-depth trade-off relation which helps us to construct a family of circuits. The designed circuits have flexible depth and number of qubits, but the total number of quantum gates is always $\mathcal{O}(mn)$. These proposals can be used as an important subroutine for estimating entanglement spectroscopy \cite{Yirka2021Qubit}, quantum metrology \cite{Yamamoto2022Error} and calculating the nonlinear function of density matrix \cite{Ekert2002Direct}. Moreover, we have applied the UMT to achieve the exponential error suppression for quantum error mitigation and numerically find that our circuits still work in the noise situation of the global depolarizing channel. Notice that the recent work \cite{Zhou2022A} estimated the MT with randomized measurement and further reduced the overhead of qubits \cite{Elben2022The}. However, the number of depth remains the same as the qubit-optimal method \cite{Ekert2002Direct}. Our algorithm gives also an alternative of the quantum parts in \cite{Zhou2022A}.

There are two proposals involving dual-state purification \cite{Huo2022Dual,Cai2021Resource}, in which only the single ancillary qubit and the implementation of the dual channel are required. The related framework utilizes qubit reset technique to reduce the number of qubits \cite{Cai2021Resource} compared with our circuit for $s=1$. However, the circuit depth is still $\mathcal{O}(m)$. Our family of circuits provides alternatives to reduce the circuit depth under the consume of the number of qubits. In general, depth is a more important index than qubit overhead for a quantum circuit. It is also worth to remark that our approaches utilize ancillary system to estimate $\textrm{Tr}(O\rho^{m})$. For $m=2$, a well-known destructive SWAP test \cite{Escartin2013SWAP,Cincio2018Learning} achieves the same goal without ancillary qubit and, at the same time, also reduces the circuit depth to a constant. However, one requires to measure multiple qubits which may increase the measurement overhead.

Notably, we have dealt with the problem for arbitrary positive integer $m$. Estimating a nonlinear function of quantum states is also of fundamental and practical interests. For instance, the fidelity involves a square root of a quantum state \cite{Nielsen2000Quantum}, while the Tsallis entropies are defined by $S^{T}_{\alpha}(\rho)=\frac{1}{1-\alpha}[\textrm{Tr}(\rho^{\alpha})-1]$ for $\alpha\in(0,1)\cup(1,\infty)$ \cite{Tsallis1988Possible}. Although direct estimation is hard on a quantum computer, a hybrid quantum-classical framework \cite{Tan2021Variational} makes the computation plausible by combining quantum state learning \cite{Lee2018Learning} and approximating fractional powers. The core idea is to minimize the quantum purity which involves an estimation of MT. Thus, our proposals can be generalized to calculate a nonlinear function of quantum states in a roundabout way.

Several interesting issues should further be investigated in the near future. The first one is to simulate the effects of different types of noise in the circuit implementation of estimating $\textrm{Tr}(O\rho^{m})$ such as the amplitude damping and phase damping channels. It would be also interesting to explore circuit structures that reduce the width and depth simultaneously. However, for the calculation of MT estimation, the qubit-depth trade-off shows that the circuit depth and width are complementary computational resources. In particular, a reduction in circuit depth is often accompanied by an increase in width, and vice versa. Current available quantum computation devices often have small size in number of qubits and circuit depth. Thus, from the view of computational resources the circuit depth and width should be reduced as much as possible in quantum algorithm design. Due to the friendly decomposition of the cyclic permutation, the circuit depth in our algorithms is reduced coincidentally, although the number of ancilla qubits is increased in a control range. Our results may highlight further investigations on the depth reduction for general quantum circuits.

\bigskip

{\sf Acknowledgements:} This work is supported by the National Natural Science Foundation of China (NSFC) under Grants 12075159 and 12171044; Beijing Natural Science Foundation (Grant No. Z190005); the Academician Innovation Platform of Hainan Province.

\begin{appendix}
\section{The proof of Theorem 1}\label{T1Proof}

The Theorem \ref{UMTEsti} can be proved in a way similar to the one given in \cite{Quek2022Multivariate}, by starting with the $s$-qubit GHZ states.
Suppose we have efficiently prepared an $s$-qubit GHZ state $|\mbox{GHZ}_s\rangle=\frac{1}{\sqrt{2}}(|0\rangle^{\otimes s}+|1\rangle^{\otimes s})$ by a constant-depth quantum circuit \cite{Quek2022Multivariate}. Eq. (\ref{MT}) provides a direct way to estimate MT by calculating the real part $\textrm{Re}[\mbox{Tr}(S^{(m)}\rho^{(n,m)})]$ and the imaginary part $\mbox{Im}[\textrm{Tr}(S^{(m)}\rho^{(n,m)})]$, where $\rho^{(n,m)}=\rho_{a_1}\otimes\rho_{a_2}\otimes\cdots\otimes\rho_{a_m}$ is an arrangement of $m$ states $\{\rho_1,\rho_2,\cdots,\rho_m\}$. Due to the equivalence of Propositions \ref{TradeOff1} and \ref{TradeOff2}, here we only complete the proof according to the circuits from Proposition \ref{TradeOff1}.

UMT estimation implements $(m-1)n=\mathcal{O}(mn)$ controlled SWAP on an initial state $\Phi_0=|\mbox{GHZ}_s\rangle\langle\mbox{GHZ}_s|\otimes\rho^{(n,m)}$, giving rise to the state
\begin{align}
\Phi_1=\frac{1}{2}(\Phi_1^{(1)}+\Phi_1^{(2)}+\Phi_1^{(3)}+\Phi_1^{(4)}),
\end{align}
where
\begin{align}
&\Phi_1^{(1)}=|0\rangle\langle0|^{\otimes s}\otimes\rho^{(n,m)},\\
&\Phi_1^{(2)}=|0\rangle\langle1|^{\otimes s}\otimes\rho^{(n,m)}(S^{(m)})^{\dag},\\
&\Phi_1^{(3)}=|1\rangle\langle0|^{\otimes s}\otimes S^{(m)}\rho^{(n,m)},\\
&\Phi_1^{(4)}=|1\rangle\langle1|^{\otimes s}\otimes S^{(m)}\rho^{(n,m)}(S^{(m)})^{\dag}.
\end{align}
Next, we measure the $s$ ancillary qubits in the basis of Pauli operator $\sigma_x$ and record $\mathcal{Q}_i=0$ or $\mathcal{Q}_i=1$, $i=1,\cdots,s$, with respect to the measurement outcomes $|+\rangle=\frac{1}{\sqrt{2}}(|0\rangle+|1\rangle)$ or $|-\rangle=\frac{1}{\sqrt{2}}(|0\rangle-|1\rangle)$, respectively. After one time measurement, obtaining a bit string $(\mathcal{Q}_1\cdots\mathcal{Q}_i\cdots\mathcal{Q}_s)$, $\mathcal{Q}_i\in\{0,1\}$, means that the state of the ancillary registers has collapsed to $|\mathcal{Q}_1\rangle\otimes\cdots\otimes|\mathcal{Q}_i
\rangle\otimes\cdots\otimes|\mathcal{Q}_s\rangle$, where $|\mathcal{Q}_i\rangle=\frac{1}{\sqrt{2}}(|0\rangle+(-1)^{\mathcal{Q}_i}|1\rangle)$.

The probability of obtaining a bit string $(\mathcal{Q}_1\cdots\mathcal{Q}_i\cdots\mathcal{Q}_s)$ is given by
\begin{align}\label{Prob}
\textrm{Pr}(\mathcal{Q}_1\cdots\mathcal{Q}_s)&=\textrm{Tr}
[M_{\mathcal{Q}_1\cdots\mathcal{Q}_s}\Phi_1]\nonumber\\
&=\frac{1}{2}(\textrm{Tr}[M_{\mathcal{Q}_1\cdots
\mathcal{Q}_s}\Phi_1^{(1)}]+\textrm{Tr}
[M_{\mathcal{Q}_1\cdots\mathcal{Q}_s}\Phi_1^{(2)}]\nonumber\\
&+\textrm{Tr}[M_{\mathcal{Q}_1\cdots\mathcal{Q}_s}
\Phi_1^{(3)}]+\textrm{Tr}[M_{\mathcal{Q}_1\cdots\mathcal{Q}_s}\Phi_1^{(4)}]),\nonumber
\end{align}
where the measurement operator
\begin{align}
M_{\mathcal{Q}_1\cdots\mathcal{Q}_s}=\Big[\bigotimes_{i=1}^{s}
|\mathcal{Q}_i\rangle\langle\mathcal{Q}_i|\Big]\otimes I.
\end{align}
Hence, we have
\begin{align}
&\textrm{Tr}[M_{\mathcal{Q}_1\cdots\mathcal{Q}_s}\Phi_1^{(1)}]
=\textrm{Tr}[M_{\mathcal{Q}_1\cdots\mathcal{Q}_s}|0\rangle\langle0|^{\otimes s}\otimes\rho^{(n,m)}]=\frac{1}{2^{s}},\nonumber\\
&\textrm{Tr}[M_{\mathcal{Q}_1\cdots\mathcal{Q}_s}\Phi_1^{(4)}]
=\textrm{Tr}[M_{\mathcal{Q}_1\cdots\mathcal{Q}_s}|1\rangle\langle1|^{\otimes s}\otimes\rho^{(n,m)}]=\frac{1}{2^{s}},\nonumber\\
&\textrm{Tr}[M_{\mathcal{Q}_1\cdots\mathcal{Q}_s}\Phi_1^{(2)}]
=\frac{(-1)^{\sum_{i=1}^{s}\mathcal{Q}_i}}{2^{s}}\textrm{Tr}(\rho^{(n,m)}(S^{(m)})^{\dag}),\nonumber\\
&\textrm{Tr}[M_{\mathcal{Q}_1\cdots\mathcal{Q}_s}\Phi_1^{(3)}]
=\frac{(-1)^{\sum_{i=1}^{s}\mathcal{Q}_i}}{2^{s}}\textrm{Tr}(S^{(m)}\rho^{(n,m)}).\nonumber
\end{align}
Now the probability takes the form,
\begin{align}
\textrm{Pr}(\mathcal{Q}_1\cdots\mathcal{Q}_s)&=\textrm{Tr}
[M_{\mathcal{Q}_1\cdots\mathcal{Q}_s}\Phi_1]\nonumber\\
&=\frac{1+(-1)^{\sum_{i=1}^{s}\mathcal{Q}_i}\textrm{Re}
[\textrm{Tr}(S^{(m)}\rho^{(n,m)})]}{2^{s}}.
\end{align}
Thus, the mean of random variable $\hat{\mathcal{Q}}=(\mathcal{Q}_1,\cdots,\mathcal{Q}_s)$ is
\begin{align}
\mathbb{E}[\hat{\mathcal{Q}}]&=\sum_{\mathcal{Q}_i\in\{0,1\}}
\textrm{Pr}(\mathcal{Q}_1\cdots\mathcal{Q}_s)
(-1)^{\sum_{i=1}^{s}\mathcal{Q}_i}\nonumber\\
&=\sum_{\mathcal{Q}_i\in\{0,1\}}
\frac{(-1)^{\sum_{i=1}^{s}\mathcal{Q}_i}+\textrm{Re}
[\textrm{Tr}(S^{(m)}\rho^{(n,m)})]}{2^{s}}\nonumber\\
&=\textrm{Re}[\textrm{Tr}(S^{(m)}\rho^{(n,m)})],
\end{align}
where in the last equality we have utilized the property $\sum_{\mathcal{Q}_i\in\{0,1\}}(-1)^{\sum_{i=1}^{s}\mathcal{Q}_i}=0$. The variance of $\hat{\mathcal{Q}}$ is given by
\begin{align}
\textrm{Var}[\hat{\mathcal{Q}}]&=\mathbb{E}[\hat{\mathcal{Q}}^2]-(\mathbb{E}[\hat{\mathcal{Q}}])^2\nonumber\\
&=1-\Big(\textrm{Re}[\textrm{Tr}(S^{(m)}\rho^{(n,m)})]\Big)^2.
\end{align}

Given a sample of size $\mathcal{N}$. Consider $\mathcal{N}$ independent random variables $\hat{Q}^{(1)},\cdots,\hat{Q}^{(j)},\cdots,\hat{Q}^{(\mathcal{N})}$, where each $\hat{Q}^{(j)}=(\hat{Q}_1^{(j)},\cdots,\hat{Q}_i^{(j)},\cdots,\hat{Q}_s^{(j)})$, $\hat{Q}_i^{(j)}\in[0,1]$ for $j=1,\cdots,\mathcal{N}$ and $i=1,\cdots,s$, corresponding to one measurement results via running the above circuit one time. The mean of $\hat{\mathcal{Q}}$ is then estimated by the sample mean,
\begin{align}
\langle\hat{Q}\rangle=\sum_{j=1}^{\mathcal{N}}\frac{\hat{Q}^{(j)}}{\mathcal{N}}.
\end{align}
Let $\varepsilon\in{0,1}$ be a precision and $\delta$ a constant such that $\delta\in(0,1)$. From the Hoeffding's inequality \cite{Hoeffding1963Probability} we have
\begin{align}
\textrm{Pr}(|\langle\hat{Q}\rangle-\mathbb{E}[\hat{Q}]|\leq\varepsilon/2)\geq1-\delta,
\end{align}
and the sample complexity $\mathcal{N}=\mathcal{O}(\varepsilon^{-2}\ln(\delta^{-1}))$.

Similarly, one implements a phase gate (mapping $|0\rangle\rightarrow|0\rangle$ and $|1\rangle\rightarrow\iota|1\rangle$, $\iota=\sqrt{-1}$) on each ancillary qubit before taking measurement to obtain the estimation of imaginary part. For the random variable $\hat{R}$ we yield a similar result,
\begin{align}
\textrm{Pr}(|\langle\hat{R}\rangle-\mathbb{E}[\hat{R}]|\leq\varepsilon/2)\geq1-\delta,
\end{align}
where the expectation value $\mathbb{E}[\hat{R}]=\mbox{Im}[\mbox{Tr}(S^{(m)}\rho^{(n,m)})]$ and the variance $\mbox{Var}[\hat{R}]=1-(\mbox{Im}[\mbox{Tr}(S^{(m)}\rho^{(n,m)})])^2$. Define $\hat{V}=\{\hat{Q},\hat{R}\}$. The mean of $\hat{V}$ is an estimation of the MT and satisfies $|\langle\hat{V}\rangle-\mathbb{E}[\hat{V}]|\leq\varepsilon$, where the mean $\mathbb{E}[\hat{V}]=\mbox{Tr}(\rho_1\cdots\rho_m)$. The variance of $\hat{V}$ is given by $\mbox{Var}[\hat{V}]=1-(\mbox{Tr}(\rho_1\cdots\rho_m))^{2}$.

\section{The proof of Theorem 2}\label{T2Proof}
We observe that the numerator of Eq. (\ref{ExpeValCor}) is
\begin{align}
\mbox{Tr}(O\rho^{m})=\mbox{Tr}(\tilde{O}^{(i)}S^{(m)}\rho\otimes\cdots\otimes\rho),
\end{align}
where the observable $\tilde{O}^{(i)}=I\otimes\cdots O^{(i)}\otimes\cdots\otimes I$ and $O^{(i)}$ denotes the operator $O$ acting on the $i$th register. Let $O=\sum_{k=1}^{N_o}a_kP_k$, $a_k\in\mathbb{R}$, be an efficient decomposition of $O$, where $P_k$ are tensor products of Pauli operators. It is straightforward to show that the trace $\mbox{Tr}(O\rho^{m})$ is a linear combination of $N_o$ MT estimations,
\begin{align}
\mbox{Tr}(O\rho^{m})&=\sum_{k=1}^{N_o}a_k\mbox{Tr}(P_k\rho^{m})\\
&=\sum_{k=1}^{N_o}a_k\mbox{Tr}[P_k^{(i)}S^{(m)}(\rho\otimes\cdots\otimes\rho)].
\end{align}
The real and imaginary parts of $\mbox{Tr}(O\rho^{m})$ can be estimated separately by using similar circuit procedure. Thus, we here only consider the estimation of the real part
\begin{align}
\mbox{Re}[\mbox{Tr}(O\rho^{m})]=\sum_{k=1}^{N_o}a_k\mbox{Re}(\mbox{Tr}[P_k^{(i)}S^{(m)}(\rho\otimes\cdots\otimes\rho)]).
\end{align}

After implementing the sequences of controlled SWAP gate, we perform a controlled $P_k$ on an arbitrary register storing the state $\rho$. Theorem \ref{UMTEsti} calculates $\mbox{Re}(\mbox{Tr}[P_k^{(i)}S^{(m)}(\rho\otimes\cdots\otimes\rho)])$ by producing a random variable $\hat{W}_k$ that can be calculated by using $\mathcal{O}(\varepsilon_k^{-2}\log(\delta^{-1}))$ repetitions of a quantum circuit (designed via propositions \ref{TradeOff1} and \ref{TradeOff2}) consisting of $\mathcal{O}(mn)$ controlled SWAP gates such that
\begin{align}\label{SubProbability}
\mbox{Pr}(|\langle\hat{W}_k\rangle-\mbox{Re}(\mbox{Tr}(P_k\rho^m))|\leq\varepsilon_k)\geq1-\delta,
\end{align}
where $\varepsilon_k\in(0,1)$, $\delta\in(0,1)$ and $\langle\hat{W}_k\rangle$ is the sample mean of variable $\hat{W}_k$. The variance is $\mbox{Var}[\hat{W}_k]=1-|\textrm{Re}(\textrm{Tr}(P_k\rho^m))|^2$. Let  $\hat{\mathcal{W}}=\sum_{k=1}^{N_o}a_k\hat{W}_k$ be a new random variable. The mean of variable $\hat{\mathcal{W}}$ has the form,
\begin{align}
\mathbb{E}[\hat{\mathcal{W}}]=\sum_{k=1}^{N_o}a_k\mathbb{E}[\hat{W}_k]=\sum_{k=1}^{N_o}a_k\mbox{Re}(\mbox{Tr}(P_k\rho^m)).
\end{align}
Its variance is given by
\begin{align}\label{Variance}
\mbox{Var}[\hat{\mathcal{W}}]&=\sum_{k=1}^{N_o}|a_k|^2\mbox{Var}[\hat{W}_k]\nonumber\\
&=\sum_{k=1}^{N_o}|a_k|^2[1-\left(\mbox{Re}(\mbox{Tr}(P_k\rho^m))\right)^2]\nonumber\\
&\leq\sum_{k=1}^{N_o}|a_k|^2,
\end{align}
where the last inequality is due to the facts that $\hat{W}_1,\cdots,\hat{W}_{N_o}$ are independent and each quantity $\mbox{Var}[\hat{W}_k]\leq1$. We remark that the quantity $\left(\sum_{k=1}^{N_o}|a_k|\right)^2$ indicates the spread of data from mean $\mathbb{E}[\hat{\mathcal{W}}]$. Again, the mean $\mathbb{E}[\hat{\mathcal{W}}]$ can be calculated by repeating the procedure $\mathcal{N}_f$ times, such that $\mathbb{E}[\hat{\mathcal{W}}]\approx\langle\hat{\mathcal{W}}\rangle
=\frac{1}{\mathcal{N}_f}\sum_{l=1}^{\mathcal{N}_f}\hat{\mathcal{W}}^{(l)}$, where $\hat{\mathcal{W}}^{(l)}$ is the measurement result on the $l$-th iteration. Moreover, the error of the estimator is
\begin{align}
|\langle\hat{\mathcal{W}}\rangle-\textrm{Re}(\mbox{Tr}(O\rho^m))|
&=\left|\langle\hat{\mathcal{W}}\rangle-\mathbb{E}[\hat{\mathcal{W}}]\right|\nonumber\\
&=\left|\sum_{k=1}^{N_o}a_k\left(\langle\hat{W}_k\rangle-\textrm{Re}[\mbox{Tr}(P_k\rho^m)]\right)\right|\nonumber\\
&\leq\sum_{k=1}^{N_o}\left|a_k\right|\varepsilon_k<\varepsilon,
\end{align}
where in the last inequality we have set $\varepsilon_1=\cdots=\varepsilon_k=\varepsilon/\sum_{k=1}^{N_o}|a_k|$, and the last equation utilizes the Eq. (\ref{SubProbability}). Using the same trick, we can estimate the imaginary part.

For $k$ runs from $1$ to $N_o$, the sample complexity is
\begin{align}
\mathcal{N}_f&=\sum_{k=1}^{N_o}\frac{1}{\varepsilon_k^2}\ln\frac{1}{\delta}
=N_o\frac{(\sum_{k=1}^{N_o}|a_k|)^2}{\varepsilon^2}\ln\frac{1}{\delta}\nonumber\\
&=\mathcal{O}\Big(\frac{N_oc^2}{\varepsilon^2}\ln\frac{1}{\delta}\Big).
\end{align}
Therefore, the total number of copies of $\rho$ is $\mathcal{O}\Big(\frac{mN_oc^2}{\varepsilon^2}\ln\frac{1}{\delta}\Big)$. We set the quantity $\sum_{k=1}^{N_o}|a_k|=\mathcal{O}(c)$ bounded by a constant $c$. Back to Eq. (\ref{Variance}), the variance is also bounded since $\sum_{k=1}^{N_o}|a_k|^2\leq\left(\sum_{k=1}^{N_o}|a_k|\right)^2$.
\end{appendix}


\begin{thebibliography}{99}
\bibitem{Shor1994} P. Shor, in \textit{Symposium on Foundations of Computer Science} (IEEE, Piscataway, NJ, 1994), pp. 124-134.
\bibitem{Harrow2009Quantum} A. W. Harrow, A. Hassidim, and S. Lloyd, \href{https://doi.org/10.1103/PhysRevLett.103.150502}{Phys. Rev. Lett. \textbf{103}, 150502 (2009)}.
\bibitem{Biamonte2017Quantum} J. Biamonte, P. Wittek, N. Pancotti, P. Rebentrost, N. Wiebe, and S. Lloyd, \href{https://doi.org/10.1038/nature23474}{Nature \textbf{549}, 195 (2017).}
\bibitem{Liang2019Quantum} J.-M. Liang, S.-Q. Shen, M. Li, and L. Li, \href{https://doi.org/10.1103/PhysRevA.99.052310}{Phys. Rev. A \textbf{99}, 052310 (2019).}
\bibitem{Cerezo2021Variational} M. Cerezo, A. Arrasmith, R. Babbush, S. C. Benjamin, S. Endo, K. Fujii, J. R. McClean, K. Mitarai, X. Yuan, L. Cincio, and P. J. Coles, \href{https://doi.org/10.1038/s42254-021-00348-9}{Nat. Rev. Phys \textbf{3}, 625 (2021).}
\bibitem{Bharti2022Noisy} K. Bharti, A. Cervera-Lierta, T. H. Kyaw, T. Haug, S. Alperin-Lea, A. Anand, M. Degroote, H. Heimonen, J. S. Kottmann, T. Menke, W.-K. Mok, S. Sim, L.-C. Kwek, and A. Aspuru-Guzik, \href{https://doi.org/10.1103/RevModPhys.94.015004}{Rev. Mod. Phys. \textbf{94}, 015004 (2022).}
\bibitem{Devitt2013Quantum} S. J Devitt, W. J. Munro and K. Nemoto, \textit{Quantum error correction for beginners}, \href{https://doi.org/10.1088/0034-4885/76/7/076001}{Rep. Prog. Phys. \textbf{76}, 076001 (2013).}
\bibitem{Lidar2013Quantum} D. Lidar and T. Brun, \emph{Quantum Error Correction} (Cambridge University Press, Cambridge, England, 2013).
\bibitem{Preskill2018Quantum} J. Preskill, \href{https://doi.org/10.22331/q-2018-08-06-79}{Quantum \textbf{2}, 79 (2018).}
\bibitem{Endo2021Hybrid} S. Endo, Z. Cai, S. C. Benjamin, and X. Yuan, \href{https://doi.org/10.7566/JPSJ.90.032001}{J. Phys. Soc. Jpn. \textbf{90}, 032001 (2021)}
\bibitem{Zhang2021Variational} S.-X. Zhang, Z.-Q. Wan, C.-Y. Hsieh, H. Yao, and S. Zhang, \href{https://arxiv.org/abs/2112.10380}{arXiv:2112.10380.}
\bibitem{Bultrini2021Unifying} D. Bultrini, M. H. Gordon, P. Czarnik, A. Arrasmith, P. J. Coles, and L. Cincio, \href{https://arxiv.org/abs/2107.13470}{arXiv:2107.13470.}
\bibitem{Yang2021Accelerated} Y. Yang, B.-N. Lu, and Y. Li, \href{https://doi.org/10.1103/PRXQuantum.2.040361}{PRX Quantum \textbf{2}, 040361 (2021).}
\bibitem{Cai2021A} Z. Cai, \href{https://arxiv.org/abs/2110.05389}{arXiv:2110.05389.}
\bibitem{Takagi2022Universal} R. Takagi, H. Tajima, and M. Gu, \href{https://arxiv.org/abs/2208.09178}{arXiv:2208.09178.}
\bibitem{Huo2022Dual} M. Huo and Y. Li, \href{https://doi.org/10.1103/PhysRevA.105.022427}{Phys. Rev. A \textbf{105}, 022427 (2022).}
\bibitem{Czarnik2022Improving} P. Czarnik, M. McKerns, A. T. Sornborger, and L. Cincio, \href{https://arxiv.org/abs/2204.07109}{arXiv:2204.07109.}
\bibitem{Cao2022NISQ} N. Cao, J. Lin, D. Kribs, Y.-T. Poon, B. Zeng, and R. Laflamme, \href{https://arxiv.org/abs/2111.02345}{arXiv:2111.02345.}
\bibitem{Liang2022Improved} J.-M. Liang, Q.-Q. Lv, S.-Q. Shen, M. Li, Z.-X. Wang, and S.-M. Fei, \href{https://doi.org/10.1002/qute.202200090}{Adv. Quantum Technol. 2200090 (2022)}.
\bibitem{McArdle2019Variational} S. McArdle, T. Jones, S. Endo, Ying Li, S. C. Benjamin, and X. Yuan, \href{https://doi.org/10.1038/s41534-019-0187-2}{npj Quantum Inf. \textbf{5}, 75 (2019)}.
\bibitem{Temme2017Error} K. Temme, S. Bravyi, and J. M. Gambetta, \href{https://doi.org/10.1103/PhysRevLett.119.180509}{Phys. Rev. Lett. \textbf{119}, 180509 (2017).}
\bibitem{Li2017Efficient} Y. Li and S. C. Benjamin, \href{https://doi.org/10.1103/PhysRevX.7.021050}{Phys. Rev. X \textbf{7}, 021050 (2017).}
\bibitem{Endo2018Practical} S. Endo, S. C. Benjamin, and Y. Li, \href{https://doi.org/10.1103/PhysRevX.8.031027}{Phys. Rev. X \textbf{8}, 031027 (2018).}
\bibitem{Yoshioka2022Generalized} N. Yoshioka, H. Hakoshima, Y. Matsuzaki, Y. Tokunaga, Y. Suzuki, and S. Endo, \href{https://doi.org/10.1103/PhysRevLett.129.020502}{Phys. Rev. Lett. \textbf{129}, 020502 (2022).}
\bibitem{Koczor2021Exponential} B. Koczor, \href{https://doi.org/10.1103/PhysRevX.11.031057}{Phys. Rev. X \textbf{11}, 031057 (2021).}
\bibitem{Huggins2021Virtual} W. J. Huggins, S. McArdle, T. E. O'Brien, J. Lee, N. C. Rubin, S. Boixo, K. B. Whaley, R. Babbush, and J. R. McClean, \href{https://doi.org/10.1103/PhysRevX.11.041036}{Phys. Rev. X \textbf{11}, 041036 (2021).}
\bibitem{Quek2022Multivariate} Y. Quek, M. M. Wilde, and E. Kaur, \href{https://arxiv.org/abs/2206.15405}{arXiv:2206.15405.}
\bibitem{Yao2010Entanglement} H. Yao and X.-L. Qi, \href{https://doi.org/10.1103/PhysRevLett.105.080501}{Phys. Rev. Lett. \textbf{105}, 080501 (2010).}
\bibitem{Johri2017Entanglement} S. Johri, D. S. Steiger, and M. Troyer, \href{https://doi.org/10.1103/PhysRevB.96.195136}{Phys. Rev. B \textbf{96}, 195136 (2017).}
\bibitem{Ekert2002Direct} A. K. Ekert, C. M. Alves, D. K. L. Oi, M. Horodecki, P. Horodecki, and L. C. Kwek, \href{https://doi.org/10.1103/PhysRevLett.88.217901}{Phys. Rev. Lett. \textrm{88}, 217901 (2002).}
\bibitem{Czarnik2021Qubit} P. Czarnik, A. Arrasmith, L. Cincio, and P. J. Coles, \href{https://arxiv.org/abs/2102.06056}{arXiv:2102.06056.}
\bibitem{Beckey2021Computable} J. L. Beckey, N. Gigena, P. J. Coles, and M. Cerezo, \href{https://doi.org/10.1103/PhysRevLett.127.140501}{Phys. Rev. Lett. \textbf{127}, 140501 (2021).}
\bibitem{Cai2021Resource} Z. Cai, \href{https://arxiv.org/abs/2107.07279}{arXiv:2107.07279.}
\bibitem{Hoeffding1963Probability} W. Hoeffding, \href{https://doi.org/10.1080/01621459.1963.10500830}{J. Am. Stat. Assoc. 58, 13 (1963).}
\bibitem{Peruzzo2014Quantum} A. Peruzzo, J. McClean, P. Shadbolt, M.-H. Yung, X.-Q. Zhou, P. J. Love, A. Aspuru-Guzik, and J. L. O'Brien, \href{https://doi.org/10.1038/ncomms5213}{Nat. Commun. \textbf{5}, 4213 (2014).}
\bibitem{Roggero2020Short} A. Roggero and A. Baroni, \href{https://doi.org/10.1103/PhysRevA.101.022328}{Phys. Rev, A \textbf{101}, 022328 (2020).}
\bibitem{Small2010Expansions} C. G. Small, \emph{Expansions and Asymptotics for Statistics} (1st ed.). Chapman and Hall/CRC, 2010.
\bibitem{Yirka2021Qubit} J. Yirka and Y. Subasi, \href{https://doi.org/10.22331/q-2021-09-02-535}{Quantum \textbf{5}, 535 (2021).}
\bibitem{Yamamoto2022Error} K. Yamamoto, S. Endo, H. Hakoshima, Y. Matsuzaki, and Y. Tokunaga, \href{https://arxiv.org/abs/2112.01850}{arXiv:2112.01850.}
\bibitem{Zhou2022A} Y. Zhou and Z. Liu, \href{https://arxiv.org/abs/2208.08416}{arXiv:2208.08416.}
\bibitem{Elben2022The} A. Elben, S. T. Flammia, H.-Y. Huang, R. Kueng, J. Preskill, B. Vermersch, and P. Zoller, \href{https://doi.org/10.1038/s42254-022-00535-2}{Nat. Rev. Phys. (2022).}
\bibitem{Escartin2013SWAP} J. C. Garcia-Escartin and P. Chamorro-Posada, \href{https://doi.org/10.1103/PhysRevA.87.052330}{Phys. Rev. A \textbf{87}, 052330 (2013)}.
\bibitem{Cincio2018Learning} L. Cincio, Y. Subaşı, A. T. Sornborger, and P. J. Coles, \href{https://doi.org/10.1088/1367-2630/aae94a}{New J. Phys. \textbf{20} 113022 (2018)}.
\bibitem{Nielsen2000Quantum} M. A. Nielsen and I. L. Chuang, \emph{Quantum Computation and Quantum Information} (Cambridge Univ. Press, 2000).
\bibitem{Tsallis1988Possible}  C. Tsallis, \href{https://doi.org/10.1007/BF01016429}{J. Stat. Phys. \textbf{52}, 479 (1988).}
\bibitem{Tan2021Variational} K. C. Tan and T. Volkoff, \href{https://doi.org/10.1103/PhysRevResearch.3.033251}{Phys. Rev. Research \textbf{3}, 033251 (2021).}
\bibitem{Lee2018Learning} S. M. Lee, J. Lee, and J. Bang, \href{https://doi.org/10.1103/PhysRevA.98.052302}{Phys. Rev. A \textbf{98}, 052302 (2018).}
\end{thebibliography}
\end{document}